\documentstyle[emulateapj]{article}

\lefthead{Taniguchi and Anabuki}
\righthead{THE ELECTRON SCATTERING REGION IN SEYFERT NUCLEI}

\begin{document}

\title{THE ELECTRON SCATTERING REGION IN SEYFERT NUCLEI}

\author{Yoshiaki Taniguchi$^1$, \& Naohisa Anabuki$^{2, 3}$}
\affil{$^1$Astronomical Institute, Graduate School of Science,
       Tohoku University, Aramaki, Aoba, Sendai 980-8578, Japan\\
       $^2$Department of Astronomy, Graduate School of Science, 
       The University of Tokyo, 7-3-1 Hongo, Bunkyo, 
       Tokyo 113-0033, Japan\\
       $^3$ISAS, 3-1-1 Yoshinodai, Sagamihara, Kanagawa
       229-8510, Japan}

\begin{abstract}
The electron scattering region (ESR) is one of important ingredients
in Seyfert nuclei because it makes possible to observe the 
hidden broad line region (hereafter HBLR) in some type 2 
Seyfert nuclei (hereafter S2s). However, little is known about
its physical and geometrical properties.
Using the number ratio of S2s with and without HBLR,
we investigate statistically where the ESR is in Seyfert nuclei.
Our analysis suggests that the ESR is located at radius between
$\sim$ 0.01 pc and $\sim$ 0.1 pc from the central engine. 
We also discuss a possible origin of the ESR briefly.
\end{abstract}


%----------------------------------------------------------------------------

\section{INTRODUCTION}

The current unified model of Seyfert nuclei has introduced a dusty
torus which surrounds the central engine. Since the torus is considered 
to be optically very thick, the visibility of the central engine is 
significantly affected by the viewing angle toward Seyfert nuclei
[Antonucci \& Miller 1985; Krolik \& Begelman 1988; 
Heisler, Lumsden, \& Bailey 1997 (hereafter HLB97);
see for a review Antonucci 1993]; i.e., if we observe Seyfert nuclei
from favored (unfavored) viewing angles, we can (cannot) see the central
engine as well as the broad line region (BLR). This explains why there 
are two types (type 1 and type 2) of Seyfert nuclei (hereafter S1s 
and S2s); the orientation of S2s is such that the BLR is obscured 
from the line of sight.
The most important observational evidence for this unified model
is the detection of a hidden BLR (HBLR) in the nearby S2, NGC 1068
in optical spectropolarimetry (Antonucci \& Miller 1985). 
Subsequent optical spectropolarimetric observations have detected the HBLR
in more than a dozen of S2s [Miller \& Goodrich 1990 (hereafter MG90);
Tran, Miller, \& Kay 1992; Kay 1994; Tran 1995a, 1995b, 1995c;
Inglis et al. 1993; Hines \& Wills 1993; Young et al. 1996; HLB97].
Although the discovery of the HBLR in the S2s has led to 
the unified model, the HBLR has been detected in only 20\%
of the observed S2s (Kay 1994; Tran 1995a). Why do some S2s
have the HBLR while the others have no HBLR ?
This raises a question whether or not all S2s are in fact S1s.

The visibility of the HBLR in S2s may be controlled by the location of
the electron scattering region (hereafter the ESR) because the light 
from the BLR is scattered by free electrons in the ESR and thus comes 
along our line of sight as the HBLR (Antonucci \& Miller 1985; MG90).
Therefore, the ESR is one of important ingredients in Seyfert nuclei.
Although the nature of the ESR is not clear,
the ESR is spatially resolved in the nearby S2, NGC 1068;
gaseous blobs with several tens pc are observed up to a radius of
$\simeq$ 300 pc (Capetti et al. 1995b; see also Kishimoto 1999).
Recently, HLB97 made a systematic search for the HBLR
for an infrared-selected sample of 16 S2s. They found that the S2s
without HBLR are more reddened in the optical and tend to have
stepper infrared colors between 25 $\mu$m and 60 $\mu$m.
These findings mean that the S2s
without HBLR are viewed from an almost edge-on view toward the dusty
torus and thus the ESR is also hidden by the obscuring material.
On the other hand, the S2s with HBLR are not so reddened [e.g.,
$E(B-V) \leq 1$] and tend to have flatter 25 $\mu$m-to-60 $\mu$m
colors, suggesting that the S2s with HBLR are viewed at an angle
intermediate between S1s and S2s without HBLR.
In summary, HLB97 adopted the following
characteristic radii; $< 1$ kpc for the NLR, $< 300$ pc
for the ESR, and $< 1$ pc for the BLR.
The characteristic radius of ESR 
is much larger than that of a typical
compact dusty torus, $<$ 1 pc (e.g., Taniguchi \& Murayama 1998
and references therein).
If all S2s have such extended ESRs and
the electron scattering is optically thin, one would find
high polarizations above 50\% in many S2s (MG90; cf. Wolf \&
Henning 1999). However, the
observed polarizations are usually less than 30\% even if
the correction for effect of unpolarized featureless continuum
is made (e.g., Tran 1995c).
Although the spatially extended ESR is observed in NGC 1068,
there seems no reason to regard NGC 1068 as a standard S2.
There may be a possibility for the majority of S2s that 
the ESR is hidden by a compact, nuclear dusty torus whose size is $< 1$ pc. 
In this Letter, based on this
idea, we estimate the typical radial distance of the ESR
in Seyfert nuclei and then discuss a possible origin of the ESR.
Hereafter, we will refer an S2 with
and without the HBLR as S2$^+$ and S2$^-$, respectively.

\section{WHERE IS THE ELECTRON SCATTERING REGION IN SEYFERT NUCLEI ?}

According to HLB97, it is strongly suggested that 
the ESR in the S2$^{-}$s is hidden by the obscuring 
material which is either a dusty torus or more extended dusty matter.
Although some Seyfert galaxies such as NGC 1068 have 
dense gaseous matter in the circumnuclear regions
(Tacconi, L. 1998, private communication; Kohno et al. 1996),
we assume for simplicity that the obscuring matter is the
dusty torus itself in all S2s.
In Figure 1, we show the geometrical relationship between the ESR 
and the torus adopted in this Letter. Although
several dusty torus models for active galactic nuclei (AGNs)
have been proposed 
[Efstathiou \& Rowan-Robinson 1990; Pier \& Krolik 1992, 1993
(hereafter PK92 and PK93, respectively); Granato \& Danese 1994;
Granato, Danese, \& Franceschini 1997),
models of PK92 are quite simple and thus are very useful
in investigating statistical properties of dusty tori (PK93;
Murayama, Mouri, \& Taniguchi 1999).
Therefore, we adopt the simple torus models of 
PK92 who studied the thermally reradiated infrared spectra of 
the compact dusty tori surrounding the central engine of AGNs
by using a two-dimensional radiative transfer
algorithm. The torus is a cylinder of dust with a uniform density,
characterized by the inner radius ($a$), the outer radius ($b$),
and the full height ($h$). A semi-opening angle of the
torus is thus given  as $\theta_{\rm open}$ =  tan$^{-1}$($2a/h$).

Given the current unified model of Seyfert nuclei,
the number ratio between S1s and S2s can be used to estimate the
opening angle of dusty tori from a statistical point of view
(e.g., Osterbrock \& Shaw 1988; MG90; Lawrence 1991); i.e.,

\begin{equation}
{{N({\rm S1})} \over {N({\rm S1}) + N({\rm S2})}} = 
\Omega / 4 \pi = 1 - {\rm cos} ~\theta_{\rm open}
\end{equation}

\noindent where $N({\rm S1})$ and $N({\rm S2})$ are the observed numbers
of S1s and S2s, respectively, $\Omega$ is 
the solid angle subtended at the source not covered by dust,
and $\theta_{\rm open}$ is the semi-opening angle of the torus.
Adopting the observed number ratios, $N({\rm S1})/N({\rm S2}) = 0.125$
(Osterbrock \& Shaw 1988), 0.20 (Salzer et al. 1989), 0.435
(Huchra \& Burg 1992),  we obtain $\theta_{\rm open} \simeq 27^\circ$,
34$^\circ$, and 46$^\circ$, respectively.
The semi-opening angle derived here is basically quite similar to that of
the bi-conical narrow line regions (NLRs) if the NLRs are collimated by
the dusty tori (e.g., Murayama et al. 1999).
Optical emission-line imaging surveys made by 
Pogge (1989), Wilson \& Tsvetanov (1994), and Schmitt \& Kinney (1996)
show that a typical semi-opening angle of dusty tori is 
$\theta_{\rm open} \simeq 30\arcdeg$.
Accordingly, it seems reasonable to adopt 
$\theta_{\rm open} \simeq 30^\circ$ 
(i.e., $N_{\rm S1}/N_{\rm S2} \simeq 0.15$). This means that
the viewing angle toward S2s lies in a range between $30^\circ$
and $90^\circ$. 

Now let us assume that 
the S2$^+$s are viewed at an angle intermediate
between S1s and S2$^-$s (HLB97).
Using the observed number ratio between S2$^+$ and S2$^-$,
$N({\rm S2}^+)/N({\rm S2}^-)$, together with the observed number of S1s
we estimate a semi angle which distinguishes S2$^+$ from
S2$^-$ ($\theta_{\rm ESR}$; see Figure 1); i.e.,

\begin{equation}
{{N({\rm S1}) + N({\rm S2}^+)} \over {N({\rm S1}) + N({\rm S2})}} =
1 - {\rm cos} ~\theta_{\rm ESR}.
\end{equation}

\noindent
There are two surveys for the HBLR in S2s\footnote{Note that 
we do not include the ultraluminous (the infrared luminosity
exceeds $10^{12} L_\odot$; see Sanders \& Mirabel 1996)
type 2 AGNs with the HBLR
(see Inglis et al. 1993; Hines \& Wills 1993; Young et al. 1996)
in our analysis.}. 
The first survey has been promoted by using the Shane 3 m
telescope at Lick Observatory. 
Ten S2s are known to have the HBLR among 50 S2s
studied so far by this survey
(Kay 1994; Tran 1995a, 1995c  and references therein).
Although the S2s are chosen to be brighter than $V$ = 16 and have
declinations north of $-25^\circ$, the sample of 50 S2s is not a 
statistically complete one. It is also noted that 
the spectropolarimetry was done in the blue where the 
signal-to-noise ratio is lower than at H$\alpha$, making detections of broad 
H$\beta$ difficult. Therefore, the detection rate in the Lick survey
is regarded as a lower limit.
The Lick survey gives an observed number ratio between S2$^+$ and S2$^-$,
$N({\rm S2}^+)/N({\rm S2}^-) = 0.25 (=10/40)$.
HLB97 made a spectropolarimetric survey for 
an infrared-selected sample of sixteen S2s at declinations south of 
$+20^\circ$, with the Galactic latitude $\vert b \vert < 20^\circ$,
and with high-quality detections at 12, 25, 60, and 100 $\mu$m,
a 60-$\mu$m flux of $f_{60} >$ 5 Jy, a far-infrared (FIR) luminosity 
cutoff of log($L_{\rm FIR}/L_\odot) >$ 9.85, and a FIR flux ratio 
of $f_{60} / f_{25} < 8.5$ to exclude normal spiral galaxies dominated
by the cool disk component.
They found that seven S2s have the HBLR, giving a ratio of
$N({\rm S2}^+)/N({\rm S2}^-) = 0.78 (=7/9)$. This ratio is
higher by a factor of 3 than that obtained in the Lick survey.
Although the HLB97 sample seems to be more statistically 
complete than the Lick one, it includes two ultraluminous infrared
galaxies [IRAS 05189$-$2524 (e.g., Sanders et al. 1988), and 
IRAS 19254$-$7245 (Mirabel, Lutz, \& Maza 1991)]
and one radio-loud AGN (PKS 2048$-$57 = IC 5063; see, e.g., Simpson,
Ward, \& Kotilainen 1994). Two (IRAS 05189$-$2524 and IC 5063)
have the HBLR while the other has no HBLR. If we exclude these three
galaxies, the number ratio could be 0.63 (5/8).
Furthermore, one of their selection criteria, 
the high-quality detection at 12 $\mu$m may reduce the number of
S2$^-$ since its mid-infrared emission is usually weak
because of the heavy extinction by the torus itself (e.g., PK92).
Therefore, we consider that the number ratio derived from 
the HLB97 sample may be overestimated although the qualitative conclusion
in HLB97 is not affected by this possible overestimation.

Using equation (2), we obtain 
$\theta_{\rm ESR} \simeq 50^\circ$ for the Lick survey
and $\theta_{\rm ESR} \simeq 80^\circ$ for the HLB97 survey.
As mentioned above, the number ratio between S2$^+$ and S2$^-$
derived from HLB97 survey may be overestimated;
note that $\theta_{\rm ESR} \simeq 65^\circ$ if 
$N({\rm S2}^+)/N({\rm S2}^-) \simeq 0.5$.
Although the Lick S2 sample is not statistically complete,
we adopt $\theta_{\rm ESR} \simeq 50^\circ$ in later discussion
keeping the result based on the HLB97 sample in mind.
Then one may summarize that 
the viewing angle toward the S2$^+$s lies in the
range between $30^\circ$ and $50^\circ$ while that toward the
S2$^-$s in the range between $50^\circ$ and $90^\circ$.

Now we can estimate the maximum radial distance of the ESR,
$r_{\rm ESR}$, given the typical dimension of the dusty torus of 
Seyfert nuclei. Using both dusty torus models of PK92 and 
high-resolution VLBI imaging data of water vapor maser emission
at 22 GHz (Miyoshi et al. 1995; Greenhill et al. 1996, 1997),
Taniguchi \& Murayama (1998) have shown that the observed
inner radii of the water vapor molecular clouds are almost 
identical to those of dusty tori. Therefore, the inner radii of 
dusty tori in Seyfert nuclei range from $a \simeq 0.1$ pc (NGC 4258;
Miyoshi et al. 1995) to $a \simeq 0.5$ pc (NGC 1068; Greenhill et al.
1996). For the most probable dusty torus model for Seyfert nuclei
(PK92, PK93; Murayama et al. 1999), the full height of the torus
is $h \simeq 3 a$ and thus $h \sim 0.3$ -- 1.5 pc.
The relation between $r_{\rm ESR}$ and $h$ is given by

\begin{equation}
{r_{\rm ESR} \over {h/2}} = 
{{{\rm tan} ~ \theta_{\rm ESR} - {\rm tan} ~ \theta_{\rm open}} \over 
{{\rm tan} ~ \theta_{\rm ESR}}} 
\simeq 0.52.
\end{equation}

\noindent We thus obtain $r_{\rm ESR} \simeq 0.26 h
\simeq 0.26 (h/{\rm 1 ~ pc})$ pc.
Therefore, it is suggested that the ESR is located at radial distance
from the central engine, $r_{\rm BLR} \leq r \leq r_{\rm ESR}$
where $r_{\rm BLR}$ is the typical radial distance of the BLR;
$r_{\rm BLR} \sim$ 0.01 pc (e.g., Peterson 1993). 
Remember that this argument is basically valid only for a case that
the majority of Seyfert nuclei have a compact dusty torus such as those
studied by PK92. It is, however, known that some Seyfert galaxies
have more spatially extended obscuring clouds (e.g., Kohno et al. 1996).
It will be necessary to investigate which type of tori (i.e.,
compact or extended) is typical for Seyfert nuclei.

Finally, it is worthwhile noting that the viewing angle estimated for
the S2$^+$s is almost consistent with the lower polarizations,
i.e., $P < 30$\% (MG90; Tran 1995c),
even after correcting for probable dilution by the contamination of 
{\it unpolarized} featureless continuum emission to the observed
spectra  (Tran 1995c;  see also Heckman et al. 1995; Cid Fernandes 
\& Terlevich 1995; Dopita et al. 1998).
If the electron scattering is optically thin, as one expects it to be
(cf. Nishiura \& Taniguchi 1998),
polarizations above 50\% would be observed in many S2$^+$s (e.g., MG90;
cf. Wolf \& Henning 1999).
It is known that the polarization due to electron scattering 
depends strongly on the scattering angle ($\theta_{\rm scat}$); 
e.g., $P \simeq$ 80\% for $\theta_{\rm scat} \simeq 90^\circ$,
$P \simeq$ 50\% for $\theta_{\rm scat} \simeq 120^\circ$
(or $60^\circ$), and
$P \simeq$ 10\% -- 15\% for $\theta_{\rm scat} \simeq 150^\circ$
(or $30^\circ$) (Kishimoto 1996).
Therefore, as suggested in this Letter, if the viewing angle toward
the S2$^+$s lies in the range between 30$^\circ$ and 50$^\circ$,
the intrinsic polarizations of S2$^+$ are expected to be $\simeq$ 
10 -- 30\% at most, being consistent with the observations.
If $r_{\rm ESR} \geq h/2$, we may not explain the intrinsic
lower polarizations in the S2$^+$s. This reinforces that our assumption 
is reasonable; i.e.,
the radial distance of ESR is shorter than that of the half height
of dusty tori (Figure 1). 

\section{POSSIBLE ORIGIN OF THE ELECTRON SCATTERING REGION}

In previous section, we have found that $r_{\rm ESR}$
is of the order of 0.1 pc. In this section we consider possible
origins of free electrons in the ESR.
There may be two alternative ideas for the origin; 
1) free electrons are formed by the photoionization 
by the ionizing continuum radiation from the central engine,
or 2) free electrons are formed by the ionizing shock driven
by the radio jet.
Recent detailed morphological studies of inner regions of the NLR 
in some nearby Seyfert nuclei
have shown that the optical NLRs are associated with the radio
jet (Bower et al. 1995; Capetti et al. 1995a, 1996). These
observations have strongly suggested that the NLR associated with
the radio jet may be formed by the ionizing fast shock driven by the
radio jet rather than the photoionization (Dopita \& Sutherland
1995, 1996; Dopita et al. 1997; Bicknell et al. 1998; Falcke,
Wilson, \& Simpson 1998;
Wilson \& Raymond 1999; see also Daltabuit \& Cox 1972;
Wilson \& Ulvestad 1983; Norman \& Miley 1984).
Although it is still uncertain that the majority
of the NLR in Seyfert nuclei is formed by the ionizing shock (Laor 1998),
the spatial coincidence between the radio jets and the optical 
emission-line gas means that the ionizing shock works in part.
The present analysis has suggested that the ESR is located 
at $r <1$ pc from the nucleus. Thus the dynamical effect for the
ESR exerted by the radio jet seems much more important than that
for more distant NLR clouds. Therefore, 
we investigate the second possibility in this Letter.

Here we consider what happens when a radio jet interacts with 
the ambient gas following Norman \& Miley (1984). The jet is
characterized by the jet luminosity $L_{\rm jet}$, the jet
velocity $v_{\rm jet}$, and the solid opening angle of the jet
$\Omega_{\rm jet}$. The pressure exerted on the ambient gas
by the radio jet is estimated as

\begin{equation}
p_{\rm jet} \sim 0.01 
\left( {L_{\rm jet} \over {10^{44} ~ {\rm erg ~ s^{-1}}}} \right)
\left( {\Omega_{\rm jet}/4\pi \over {0.01}} \right)^{-1}
\left( {r_{\rm jet} \over {1 ~ {\rm pc}}} \right)^{-2}
\left( {v_{\rm jet} \over {10^5 ~ {\rm km ~ s^{-1}}}} \right)^{-1}
~ {\rm dyne ~ cm}^{-2}.
\end{equation} 

\noindent where $r_{\rm jet}$ is the radial distance of the jet.
If we assume that the ambient gas clouds can cool and reach 
pressure equilibrium in the cocoon of the jet,
we obtain a kinetic temperature of the gas

\begin{equation}
T_{\rm kin} \sim 10^4
\left( {n_{\rm e} \over {10^{10} ~ {\rm cm^{-3}}}} \right)^{-1}
\left( {L_{\rm jet} \over {10^{44} ~ {\rm erg ~ s^{-1}}}} \right)
\left( {\Omega_{\rm jet}/4\pi \over {0.01}} \right)^{-1}
\left( {r_{\rm jet} \over {1 ~ {\rm pc}}} \right)^{-2}
\left( {v_{\rm jet} \over {10^5 ~ {\rm km ~ s^{-1}}}} \right)^{-1}
~ {\rm K}
\end{equation}

\noindent where $n_{\rm e}$ is the electron density.
For typical Seyfert nuclei, $L_{\rm jet}$ is of the order of
$10^{40}$ erg s$^{-1}$ at most (e.g., Wilson, Ward, \& Haniff 1988)
and $v_{\rm jet}$ is of the order of 10$^4$ km s$^{-1}$
(e.g., Gallimore, Baum, \& O'dea 1996). 
The typical electron density in the ESR may be closer to
that of the inner surface of dusty tori ($\sim 10^8$ cm$^{-3}$;
Pier \& Voit 1995) than to that in the BLR ($\sim 10^9$ cm$^{-3}$;
e.g., Osterbrock 1989). Therefore, we adopt $n_{\rm e} = 10^8$ 
cm$^{-3}$ for the ESR (see also Nishiura \& Taniguchi 1998).
Then we obtain a typical kinetic temperature of the ESR 
in Seyfert nuclei at 
$r_{\rm jet} \sim r_{\rm ESR} \sim$ 0.1 pc,

\begin{equation}
T_{\rm kin}({\rm ESR}) \sim 10^5
\left( {n_{\rm e} \over {10^{8} ~ {\rm cm^{-3}}}} \right)^{-1}
\left( {L_{\rm jet} \over {10^{40} ~ {\rm erg ~ s^{-1}}}} \right)
\left( {\Omega_{\rm jet}/4\pi \over {0.01}} \right)^{-1}
\left( {r_{\rm ESR} \over {0.1 ~ {\rm pc}}} \right)^{-2}
\left( {v_{\rm jet} \over {10^4 ~ {\rm km ~ s^{-1}}}} \right)^{-1}
~ {\rm K}.
\end{equation}

\noindent As claimed in earlier studies (e.g., MG90), 
if the electron scattering occurs in a high-temperature gas with
$T_{\rm kin} > 10^6$ K, the line width of the HBLR will be
broadened thermally;
i.e., FWHM $\simeq 9200 (T_{\rm kin}/10^6 ~{\rm K})^{1/2}$ km s$^{-1}$.
However, the observed FWHM of the HBLR lies mostly in a range
between 3000 km s$^{-1}$ -- 6000 km s$^{-1}$ (MG90; Tran 1995a).
Therefore, the above estimate appears consistent with the observations.
According to the formulation in equation (6), if $n_{\rm e}
\propto r^{-2}$, the kinetic temperature keeps at $\sim 10^5$ K
even for more distant ionized gas clouds.
Finding the correlation between FWHM of narrow optical emission
lines and their critical densities for collisional deexcitation, 
De Robertis \& Osterbrock (1986) suggested the relation of
$n_{\rm e} \propto r^{-2}$ for S2s (see also Osterbrock 1989,
Section, 12.7). This may explain the spatially
extended ESRs in NGC 1068. However, even if there are extended ESRs,
inner ionized gas clouds will contribute more to the HBLR flux in 
polarized spectra because their covering factors are larger than
those of outer ones if the cloud size is almost similar at any $r$.

In summary, the thermal Bremsstrahlung emission from this warm gas
with $T_{\rm kin} \sim 10^5$ K can be responsible for the ionization.
It is also likely that the jet-driven ionizing shock works there
(e.g., Dopita \& Sutherland 1995; Falcke et al. 1998; Wilson \& Raymond
1999; cf. Laor 1998). The continuum radiation from these processes
may contribute to the so-called featureless continuum emission
although dilution by starlight may be important in some S2s.
Finally, we would like to mention that the ESR may play an important
role in the scattering of X-ray photons (e.g., Turner et al. 1997;
Netzer, Turner, \& George 1998; Awaki, Ueno, \& Taniguchi 1999). 

\acknowledgments

We would like to thank Hisamitsu Awaki and Takashi Murayama
for useful discussion and an anonymous referee for useful
suggestions and comments.
This work was supported in part by the Ministry of Education, Science,
Sports and Culture in Japan under Grant Nos. 10044052, and 10304013.

%----------------------------------------------------------------------
%           References
%----------------------------------------------------------------------

%----------------------------------------------------------------------
%            Figure Caption
%----------------------------------------------------------------------

\begin{figure}
\epsfysize=18.5cm \epsfbox{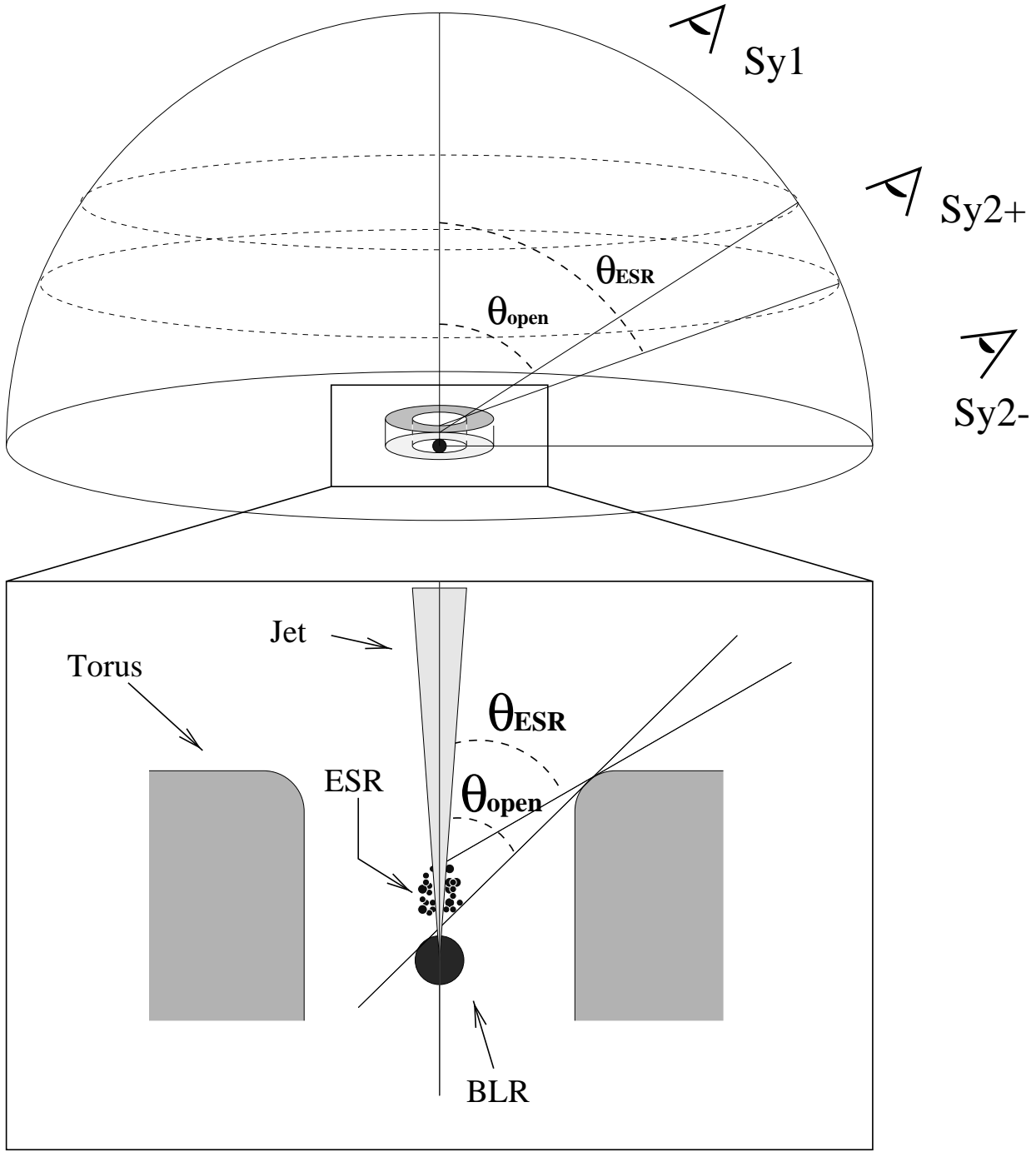}
\caption[]{
A geometrical relationship between the dusty torus and
the electron scattering region (ESR).
\label{fig1}
}
\end{figure}

%----------------------------------------------------------------------

\end{document}